\documentclass{kluwer}
\usepackage{epsfig}

\def\be{\begin{equation}}
\def\ee{\end{equation}}

\def\jcd{Christensen-Dalsgaard}
\def\apj#1.{{\it Astrophys.\ J.} {\bf #1},}

\def\ga{{\leavevmode\kern0.3em\raise.3ex\hbox{$>$}
\kern-0.8em\lower.7ex \hbox{$\sim$}\kern0.3em}}
\def\la{{\leavevmode\kern0.3em\raise.3ex\hbox{$<$}
\kern-0.8em\lower.7ex \hbox{$\sim$}\kern0.3em}}

\begin{document}
\begin{article}
\begin{opening}

\runningtitle{Solar cycle variations in the convection zone}
\runningauthor{S. Basu and H. M. Antia}

\title{POSSIBLE SOLAR CYCLE VARIATIONS IN THE \\  CONVECTION ZONE}

\author{Sarbani \surname{Basu}}
\institute{Institute for Advanced Study,
Olden Lane, Princeton NJ 08540, U. S. A., and
Astronomy Department, Yale University, P.O. Box 208101 New Haven,
CT 06520-8101 USA}
\author{H. M. \surname{Antia}}
\institute{Tata Institute of Fundamental Research,
Homi Bhabha Road, Mumbai 400005, India}
\date{\today}


\begin{abstract}
Using data from the Global Oscillations Network Group (GONG) that
covers the period from 1995 to 1998 we study the change in frequencies
of solar oscillations with solar activity. From these frequencies we
attempt to determine any possible variation in solar structure with
solar activity. We do not find any evidence of a change in the
convection zone depth or extent of overshoot below the convection
zone during the solar cycle. 
\end{abstract}

\end{opening}

\section{Introduction}
It is well known that frequencies of solar oscillations vary with time
(Elsworth et al.~1990; Libbrecht and Woodard 1990; Dziembowski et
al.~1998; Bhatnagar, Jain and Tripathy 1999; Howe, Komm and Hill 1999)
and that these variations are correlated with solar activity.
To a first approximation the
scaled frequency differences between frequencies at different phases of
the solar cycle appear to be a smooth function of frequency indicating
that the solar-cycle related changes are predominantly taking place
near the solar surface. The seat of the solar dynamo, however, is
believed to be near the base of the convection zone (hereafter CZ) and
thus one may expect some changes in this region during the solar cycle.
Monteiro, \jcd, and Thompson~(1998) have found some hint of variation
in extent of overshoot below the solar convection zone. This needs to
be confirmed by independent analysis. Thus in this work we investigate
whether these frequency differences with time imply any change in depth
of the solar convection zone or extent of overshoot with solar
activity. 

\section{Data used}
We have used monthly $m$-averaged power spectra from the Global
Oscillations Network Group (GONG) project to determine solar
oscillation frequencies. We have fitted the rotation corrected,
$m$-averaged, power spectra for each GONG month, 1 to 37. Each GONG month
extends over 36 days, with month 1 beginning on 7 May, 1995 and month
37 ending on 28 December, 1998.  To improve the
signal-to-noise ratio in the spectra used to determine the CZ depth
and the extent of overshoot below the CZ, we have summed spectra for
four months. We select 19 summed spectra covering the entire period and
solar activity range. There is some overlap in the time period covered
by these spectra.
We have also summed the spectra for GONG months 7
to 22 and 24 to 35 to improve the signal-to-noise ratio still further. The 
months 7 to 22 cover a period around minimum solar activity
extending from 9 December,
1995 to 6 July, 1997, while months 24 to 35 cover a period of high
activity from 12 August, 1997 to 17 October, 1998. These two summed
spectra provide  precise data during period of low and high activity,
thus allowing us to measure small variations in solar interior.
 
To calculate the frequencies from the $m$-averaged power spectra we fit
each mode separately using a maximum likelihood technique
(Anderson, Duvall and Jefferies~1990).
The fit is performed over a region extending midway to the next mode
on either side of the peak and includes all leaks which
occur within the fitting interval from modes of degree
$\ell-3$ to $\ell+3$, $\ell$ being the degree of the target mode.
Because of possibly incorrect splitting coefficients used in constructing
the $m$-averaged power spectra the apparent widths of the peaks may be
increased, but that is not likely to affect the fitted frequencies as the
width of each leak is treated as a separate parameter while
fitting the spectra. Further, in this paper we are interested in
the frequency shift with solar cycle so any possible systematic error
introduced in fitting the frequencies for an individual month will tend
to get cancelled while taking the frequency differences between two months.
Moreover we
have tested for possible systematic errors in fitting $m$-averaged
spectra by comparing the fitted frequencies with those computed by
the GONG project using the individual $m$ spectra
for the same period (Basu and Antia, 2000).
Systematic difference between these
two sets of frequencies is $\la10$ nHz.

In addition we also use the frequencies from Big Bear Solar Observatory (BBSO)
data for the years 1986, 1988, 1989 and 1990. These data sets cover the rising
phase of the cycle 22 (Woodard and Libbrecht 1993).
The estimated error in frequencies in these data sets is much larger
than those in GONG data.

\section{Technique}
The temperature gradient within the Sun changes from the adiabatic value 
inside the
convection zone to the radiative value below the base of the convection
zone. This sharp change in the temperature gradient introduces a bump
in the sound speed difference between two models (or between a model
and the Sun) if they have different CZ depths. This signal can be used
to determine the depth of the CZ. We use the method described by Basu
and Antia (1997) to determine the radial position of the CZ base. 

The transition of the temperature gradient from the adiabatic to
radiative values at the base of the solar convection zone also gives
rise to an oscillatory signal in frequencies of all modes which
penetrate below the base of the solar convection zone (Gough 1990) and
this signal can be used to determine the extent of overshoot below the
solar CZ (Monteiro, \jcd\ and Thompson~1994; Basu, Antia and
Narasimha~1994). This signal can be amplified by taking the fourth
differences of the frequencies as a function of the radial order $n$
enabling a more precise measurement of the amplitude, $A$, of
oscillations. This amplitude increases with the extent of overshoot and
can be calibrated against amplitudes for models with known extents of
overshoot. The frequency $\tau$ of the signal is a measure of the
acoustic depth at which the transition occurs. We use the method
described in Basu (1997) to isolate the oscillatory signal and measure
its characteristics. 

\section{Results}

\begin{figure}
\centerline{\epsfig{file=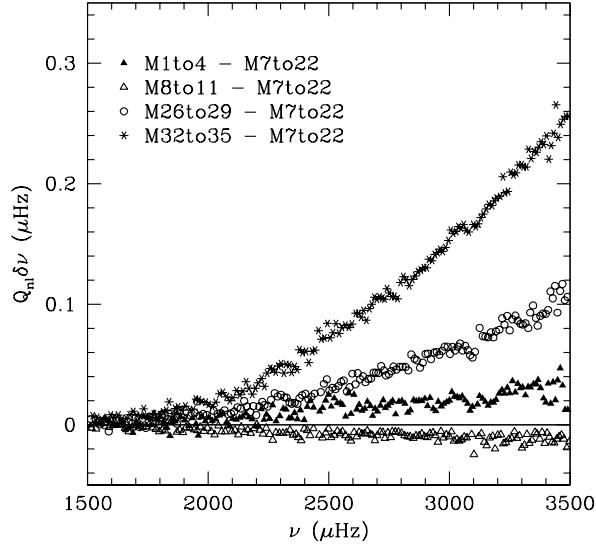,width=8.0 true cm}}
\caption{
Differences in frequencies determined from observations taken at
different times during the solar cycle.  M7to22 is the data set
for months 7 to 22, similarly M1to4, M8to11, M26to29 and M32to35 are the
data sets covering months 1--4, 8--11, 26--29 and 32--35 respectively.
The scaled frequency differences have
been averaged in groups of 10 modes to reduce scatter. 
}
\end{figure}

\begin{figure}
\centerline{\epsfig{file=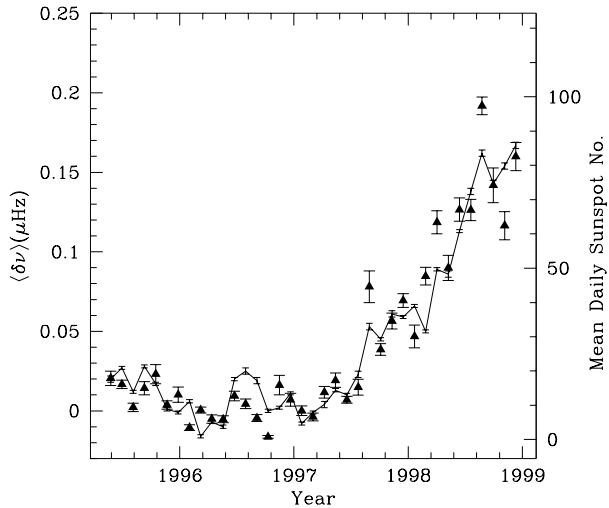,width=8.0 true cm}}
\caption{
The mean frequency shift averaged over all modes with $1.5\le\nu\le3.5$ mHz
and $5\le\ell\le100$ plotted as a function of time for the
37 GONG months is shown by the set of points which are joined by
a continuous line. The crosses show the average number of sunspots each
day during the period when the observations were taken, the scale for this
is on the right-hand axis.
}
\end{figure}

For each of these summed spectra we fit symmetric Lorentzian profiles
to calculate the frequencies using a maximum likelihood approach.
Fig.~1 shows the scaled
differences in frequencies obtained from months 1--4, 8--11, 26--29 and
32--35 and those obtained from summed spectra for months 7 to 22. The
differences have been averaged in groups of 10 modes to reduce scatter.
The frequencies are scaled by the quantity $Q_{nl}$ which is the ratio
of the mode inertia of a mode of degree $l$ order $n$ to that of a mode
of degree 0 with the same frequency as the mode of degree $l$ order
$n$ (Christensen-Dalsgaard and Berthomieu 1991). Note that to a first 
approximation the shift in frequencies is a
smooth function of frequency. We therefore may not expect much change
in the deeper layers of the Sun. The frequency difference for months
8--11 which is close to solar minimum is negative, while those from
months 1--4, which is before the minimum in cycle 22, is slightly
positive. As we approach the solar maximum the frequency difference
increases rapidly.

\begin{figure}
\centerline{\epsfig{file=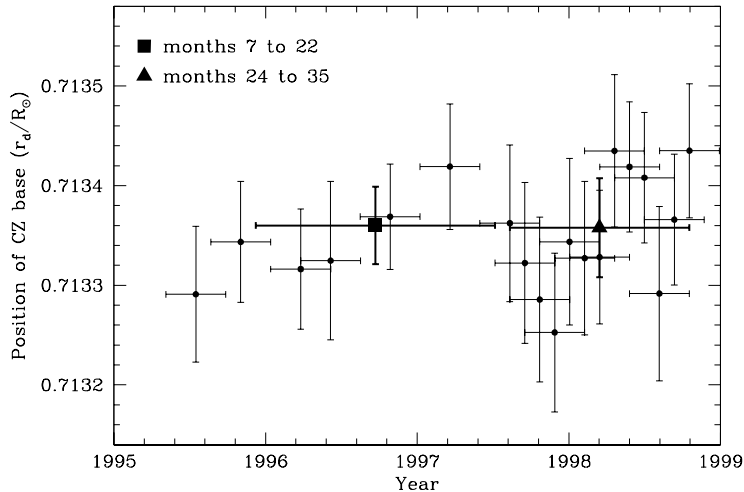,width=10.0 true cm}}
\caption{
The position of the solar CZ base plotted as 
a function of time. 
Note that the error estimates do not include
any contribution from systematic errors. These will be similar for all the
results shown. The horizontal error-bars indicate the period of time
over which the data were collected.
}
\end{figure}
\begin{figure}
\centerline{\epsfig{file=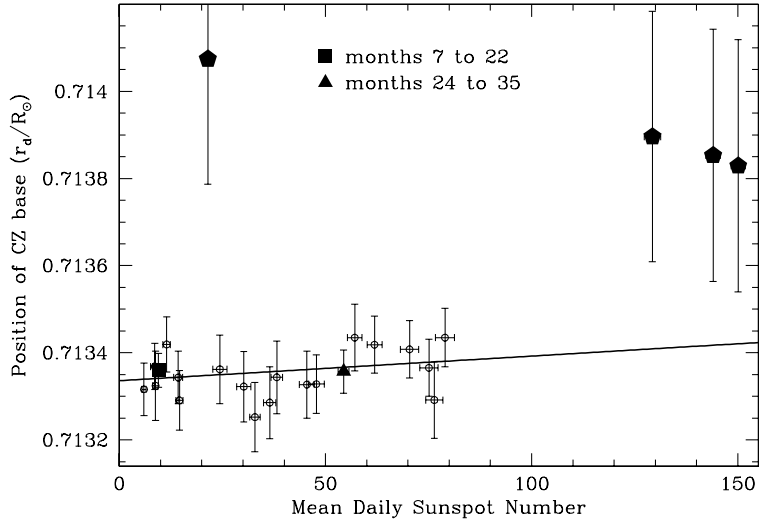,width=10.0 true cm}}
\caption{
The position of the solar CZ base plotted as 
a function of the mean daily sunspot number. The large, filled pentagons are
from the BBSO data taken during the last solar cycle.
The straight line shows the best fit to GONG points.
}
\end{figure}

The change in solar oscillation frequencies is found to be well
correlated with the activity indicators like the sunspot number. Fig.~2
shows the mean frequency shift averaged over all modes with
$1.5\le\nu\le3.5$ mHz and $5\le\ell\le100$ which are common to all data
sets, as a functions of time. This figure also shows the mean daily
sunspot number, $R_I$ during the corresponding period as obtained from
Solar Geophysical data web page (http://www.ngdc.noaa.gov/stp/stp.html)
of the US National Geophysical Data Center. The quantity $R_I$ is a
measure of solar activity.  It is quite clear that the
two quantities are correlated.  Similar figures can be drawn for other
activity indices (Bhatnagar et al.~1999).

In order to study possible temporal variation of the internal
structure of the Sun, we use the frequencies from each data set to
calculate the position of the base of the convection zone ($r_d$). The
results from GONG data are shown as a function of time in Fig.~3. There is no
systematic pattern to suggest that there are any solar cycle related
changes in the CZ depth. Fig.~4 shows the same results but plotted as a
function of the mean daily sunspot number, $R_I$. Once again there is
no clear trend visible and all points appear to be consistent with the
constant value found using summed spectra for months 7--22 or months
24--35. These values are more precise because it involves sum over more
spectra covering a larger time interval. Again there is no significant
difference between these two values. If we try to fit a straight line
through the points from GONG data in Fig.~4, then the resulting
slope is $(5.6\pm5.6)\times10^{-7}$, which is hardly significant.
More data from subsequent months with higher activity should be
able to estimate this slope more reliably.
The few results that we have from the
previous solar cycle obtained by the Big Bear Solar Observatory (BBSO)
also do not show any variation with solar activity, but the results
are systematically larger than the GONG results. This difference could
be due to systematic differences in the two data sets and may not
represent any real changes in solar interior.

\begin{figure}
\centerline{\epsfig{file=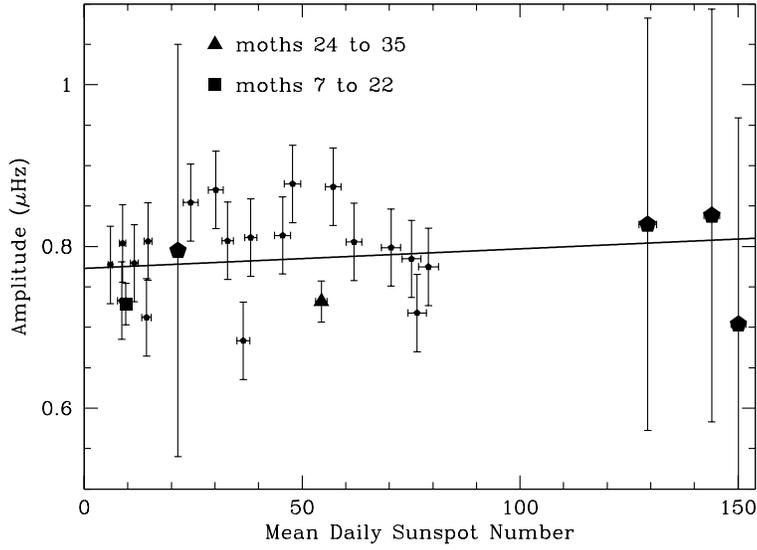,width=10.0 true cm}}
\caption{
The amplitude of the oscillatory signal in the fourth difference of
frequencies plotted as a function of the average, number of
sunspots per day during the observing period.
The large, filled pentagons are results from data obtained
by BBSO during the last solar cycle.
The straight line shows the best fit to GONG points.
}
\end{figure}

\begin{figure}
\centerline{\epsfig{file=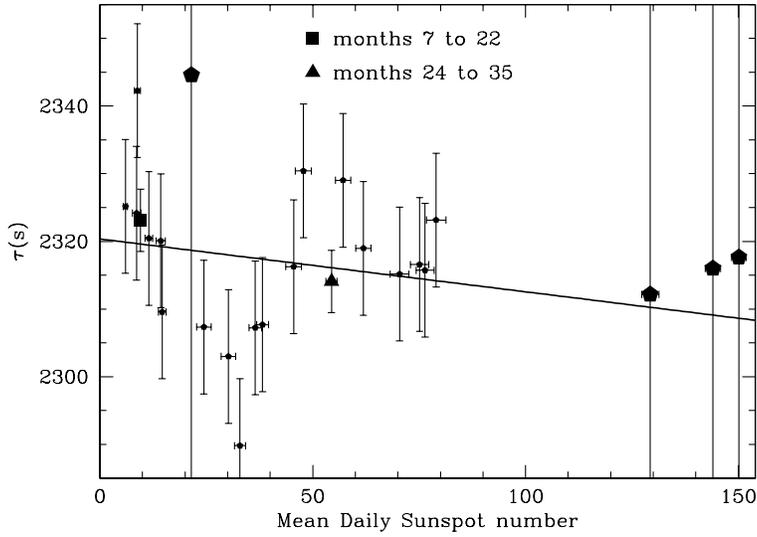,width=10.0 true cm}}
\caption{
The frequency of the oscillatory signal in the frequencies plotted as a
function of the average, daily number of sunspots. 
The large, filled pentagons are results from BBSO data taken during the
last solar cycle.
The straight line shows the best fit to GONG points.
}
\end{figure}

We also calculate the extent of overshoot below the convection zone
using all the data sets. Fig.~5 shows how the amplitude of the
oscillatory part in the fourth difference of the frequencies due to the
transition at the base of the overshoot layer,
changes as a function of the mean daily number of sunspots.
The amplitude should increase with any increase in the extent of
overshoot below the solar CZ base. The points corresponding to
4 month averaged spectra are systematically above
the mean values determined from the spectra summed over the larger time
intervals. This is because an increase in the errors of the frequencies,
in general, tends to increase the estimated amplitude of the signal
(Basu 1997).
Again
we fail to find any systematic trend to indicate solar-cycle related
changes in the extent of the overshoot layer.
The BBSO results from the previous
solar cycle are in good agreement with the GONG results.
A straight line fit through the GONG points yields a slope of
$(2.4\pm6.0)\times10^{-4}\;\mu$Hz, which is
consistent with zero.

Fig.~6 shows how the frequency of the oscillatory part in frequencies
due to the transition at the base of the overshoot layer changes
as a function of the mean daily sunspot number.
This frequency is expected to be the acoustic depth of the base of
overshoot layer. Once again we do not see any systematic variation
in the acoustic depth with time.
A straight line fit through the GONG points yields a slope of
$-0.078\pm0.077$ s, which is hardly significant.
This is consistent with the fact that we do not find any significant
variation in the CZ depth. The errors in estimated $\tau$ are much
larger than those in estimating CZ depth directly and $\tau$ may have
some uncertainty arising from those in surface layers. Some of the
variation in $\tau$ could also be due to variation in surface layers.
A variation in CZ depth by $0.0001R_\odot$ will change $\tau$ by
about 0.3 s, which is too small to be detected. Change in extent of
overshoot can also affect $\tau$ but once again the amplitude of
oscillatory signal is more sensitive to these changes.
Hence the variation, if any, in $\tau$ is likely to be due to
surface effects.

\begin{figure}
\centerline{\epsfig{file=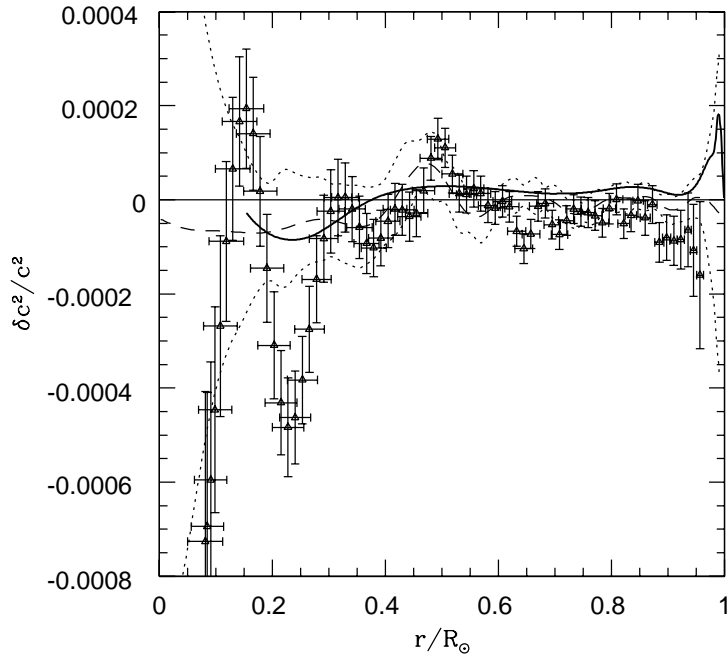,width=10.0 true cm}}
\caption{
The relative sound-speed differences that result from the difference in
frequencies between months 32--35 and 7--22. 
The points are the result of an Subtractive Optimally Localized 
Averages (SOLA) inversion. The dashed line with the dotted
lines showing the $1\sigma$ error limits are results of a
Regularized Least Squares (RLS) inversion. The thick continuous line
is a result using an asymptotic inversion. 
}
\end{figure}

Having failed to find any systematic change in the depth of the
convection zone or the extent of overshoot below the solar convection
zone, we attempt to see if there are any significant changes taking
place in the solar interior. To test this we take the frequency
difference between months 32--35 and 7--22, and apply inversion
techniques to see if there is any corresponding change in the
sound-speed profile inside the Sun. We have chosen these two sets since
solar activity changed significantly between these two intervals,
thereby causing a large change in frequencies. For other pairs of data
sets the frequency difference is smaller. Fig.~7 shows the relative
sound speed differences that result from frequency differences between
months 32--35 and 7--22 using various inversion techniques. We have
tried the asymptotic inversion technique (\jcd, Gough and Thompson
1989) as well as nonasymptotic inversion using the Regularized Least
Squares (RLS) technique as described by Antia (1996) and the
Subtractive Optimally Localized Averages (SOLA) technique as described
by Basu et al. (1996). All these inversion results agree with each
other and the resulting variation in sound speed is within the expected
errors.  These results confirm our expectation that the frequency
difference probably do not imply changes in structure deep inside the
Sun. similar results have been found for frequency difference between
other pairs of data sets. Thus it appears that the changes in frequency
with time are due to variations in surface layers, which do not affect
the deep interior significantly. 

\section{Conclusions}
The frequencies of solar oscillations show a variation which is
correlated with solar activity. We find no clear evidence of solar
cycle related change in the depth of the convection zone or the extent
of overshoot below the convection zone.
Our study has been mainly restricted to the rising phase of the cycle 23
and it would be interesting to check the variations in subsequent part
of the solar cycle. Within the period of 1995 to 1998 considered in this
study, the changes, if any, in the depth
of the CZ should be less than the error limits in our estimate, i.e.,
$0.0001R_\odot\approx 70$ km. There is no significant change in the
sound speed in deep interior. Thus it appears that solar cycle changes
in solar structure are predominantly confined to near surface regions.

\begin{acknowledgements}
This work  utilizes data obtained by the Global Oscillation
Network Group (GONG) project, managed by the National Solar Observatory, a
Division of the National Optical Astronomy Observatories, which is
operated by AURA, Inc. under a cooperative agreement with the
National Science Foundation. The data were acquired by instruments
operated by the Big Bear Solar Observatory, High Altitude Observatory,
Learmonth Solar Observatory, Udaipur Solar Observatory, Instituto de
Astrofisico de Canarias, and Cerro Tololo Inter-American Observatory.
\end{acknowledgements}

\end{article}
\end{document}